Hubble plots of the distance of stellar objects vs. recession velocity normally assume the red shift is wholly Doppler and ignore any gravitational contribution. This is unwarranted: gravity and Doppler velocity red shifts are found to be separable and contribute about equally. A recent data set, to Z=1.2, by Riess (1), was analyzed. Upon plotting distance vs. Doppler velocity, the slope of the Hubble plot increases. The Hubble plot is also curved, upwards, and this can be understood in terms of the relativistic metric changes of the space through which the light travels. On fitting the data to a simple model of a big bang of constant density, this finds the total mass of the big bang is $M=21.1 \times 10^{52}$ kg. When present actual distance is plotted vs. Doppler velocity, the plot is linear and agrees with Hubble's concept, without acceleration. Time since the big bang is longer than the 14 billion years that had been thought, 23.5 billion years. The Hubble constant hence shrinks from $H_0=71$ to $H_0=41.6$. This is an independent affirmation of a recent CMB finding of a low $H_0 \approx 35$.

## Introduction

The Hubble constant derives from Hubble's 1929 finding that light from distant nebulae is red shifted, proportional to distance. From this comes the theory that our universe arose from a single explosive event, the "big bang", with each element (galaxy) now receding from center (and from all other elements) with speed proportional to present distance. The constant of proportionality is $H_0$, the Hubble constant, whose units are km/s/mpc. $H_0$ is not really constant, of course. The inverse of $H_0$ is T, the time elapsed since the big bang and this is increasing. Upon careful examination, one finds that optically measured distance is not the geometrical distance, velocity is not measured by the red shift, Z, and the SN1a supernovae blew at different times and so a proper Hubble plot requires extrapolation of the distance to where the remnants are now at time T. This study finds ways to deal with these problems, and finds a lower $H_0$ than the present consensus value. This study also finds the total mass for the universe, and an inkling of how far out from center we are.

Astronomers measure distance optically, using well-studied light sources such as supernovae of the SN1a type. These SN1a supernovae are thought to arise when a white dwarf slowly accumulates hydrogen, and this initiates a reproducible thermonuclear fusion event when the total mass approaches 1.4 solar masses. These sources are very bright, briefly outshining the entire light output of the billions of other stars of the galaxy in which they reside. The inverse square law is then invoked for these "standard candles", and distance is calculated according to dimness. The problem with this is that optically measured distance is larger than geometric distance, and needs to be corrected for the gravitational changes in the metric of the space through which the light travels to us. Velocity, too, is a problem. It has been assumed that the red shift is wholly Doppler, although it is known that gravity also leads to a red shift. Until now, there has been found no way to extract Doppler velocity from the measured red shift, and this systematically biases the Hubble constant, upwards.

This study assumes that Hubble was right in finding that the velocity of recession of every luminous object in the night sky is proportional to distance. Before the recent high-Z SN1a data extended the range, Hubble plots tended to be linear. Linearity of such a plot implies that velocity does not change. The extension of Hubble data to far distances led to a surprising result: Hubble plots curve upwards, especially at high Z. (1) Cosmologists have long worried about whether the universe will continue to expand, making the skies darker and leaving us more and more alone in the universe. Or, could there be sufficient gravitational attraction to eventually pull all the mass back together in what has been called "the big crunch"? The curvature of Hubble plots has been interpreted to indicate that gravitational attraction is not only insufficient to slow down and reverse the recession velocity, but that the recession velocity is now increasing.

The next two figures show how astronomers display a Hubble plot, observed distance vs. red shift, Z. Data points are the red squares. Distance is in light years. The first plot appears deceptively simple, showing an apparently small deviation from the Hubble expectation (the blue line, $H_0 = 70$). It is well known that log-log plots tend to minimize any differences at the high end, and so Riess's data (1) is replotted linearly in Figure 2 to show that the curvature is not trivial.

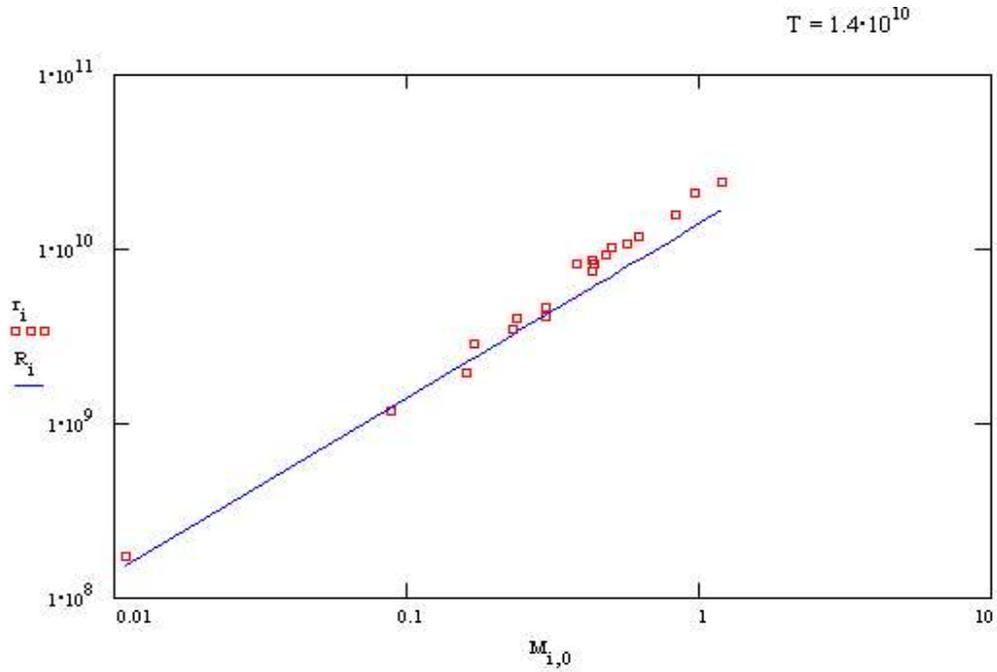

Figure 1. Log-log Hubble plot, distance vs. red shift.

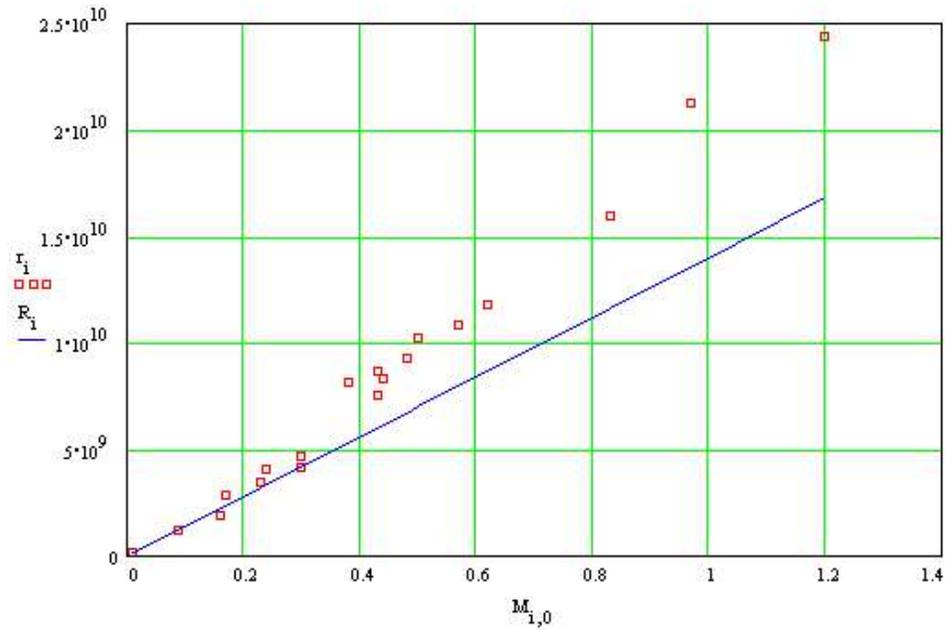

Figure 2. Linear Hubble plot, distance vs. red shift.

Why should a Hubble plot be curved? It has been suggested that recession velocities in the early universe were indeed slowing due to gravity, but this acceleration has reversed and the universe is now blowing itself apart at a faster and faster rate. (2) What could explain this? An unknown "dark energy" is assigned this task. How much, and of what kind? Of all the components of our universe that carry energy, including light, mass, etc., it is claimed that about 67% must be in the form of "dark energy" in order to account for the perceived present acceleration of the detritus of the big bang. I dislike this explanation, as it merely replaces one mystery with another.

## Geometry within the big bang.

The geometry within the big bang, which has apparently created our present universe, is mind-boggling. Nothing stands still. Every object is receding from every other, with a velocity more-or-less proportional to the present distance between the objects. When the gravitational changes of the metric are included, the very scaffold on which we navigate through the universe seems to change with time. And, there is latency. On the cosmic time scales involved, changes in the distribution of mass at one end of the universe are only effective at another place some time later due to the finite speed of light (assuming that gravity goes at the speed of light).

If we could hold everything fixed, these changes would have a chance to equilibrate, and we could then calculate the gravitational potential for the universe modeled as a sphere filled with uniform mass density (over the large scale). Given the present radius R and the total mass M, one would then find the gravitational potential at any r<R,

$$U = -(GM/2R^3)(3R^2-r^2) = -(1.5GM/R)(1-r^2/3R^2) \qquad [1]$$

U varies with "r", throughout the big bang, and this forms a basis for navigation within the big bang. In other words, our big bang is no longer a featureless plain (as the cosmological principle insists) if we can find a way to measure U. Topologically, U is bowl-shaped.

Latency affects this, because the big bang continues to expand. Standing at center, the contribution to the gravitational potential at that spot due to each successive spherical shell of mass at radius r is not that now there but is that which was there at time t-r/c. That is, the effective time is not t but is t(1−β). Latency increases the magnitude of U, since each shell at radius r contributes as though it were closer than r.  There is a second consideration, the possibility that the gravitational tug of one mass on another is diluted by the Doppler recession speed.  It seems to me reasonable to assign a diminution of the effects of gravity according to 1−β. If so, these effects cancel and we can then rely on a simple calculation of U as shown in [1].  This greatly simplifies calculations of gravitational potential within our big bang, and will be assumed so for the rest of this study.

## Distance corrections to the Hubble plot.

This study looks into the problems of correcting the Hubble plot so that it really becomes a plot of geometric distance vs. recession velocity. But first, what is meant by geometrical distance?  It is well known that gravity tends to shrink the size of a meter stick.  The distance between two fixed points can be measured in different ways, and with different results. Suppose that the two fixed points are Earth and Mars (not really fixed, but the orbits are known).  The distance from Earth to Mars can be measured in at least 3 ways.  Shapiro (3) measured it optically, using radar, and found a delay time of 248 microseconds at superior conjunction (when the path just grazed the sun) for the round trip, compared to another time when the sun's gravity was not a factor.  The gravity of the sun both increases the number of meter sticks needed to span the distance and also slows the speed of light as observed on Earth.  The one-way radar (optical) distance is increased by gravity, by 124 microseconds times c.  If one could measure the distance using meter sticks, he would find half the extra distance found by Shapiro. A third way is to use plane geometry, from a third point well away from Earth, Mars, and the sun.  With the measurements of two legs of the triangle and the subtended angle all devoid of gravitational complications, one calculates the real or Euclidean geometric distance and this is the smallest of the three measurements. It seems to me that the geometric distance, rather than the gravitationally distorted measurement of distance, is what Hubble had in mind.

Mass-metric relativity (MMR) was introduced in 2000 as an alternate to general relativity (GR). (4)  Each theory finds distortions of the measured metric of space, in the presence of gravity.  GR interprets these distortions of space as real, while MMR interprets these as changes of our standards of measurement. GR was developed by Einstein in the 1920's. MMR is a scalar theory, based on an interpretation of Pound and Rebka's 1960 Mossbauer experiment on "the weight of photons" and finds that mass increases with gravitational potential as well as with speed. Any increase of rest mass, according to quantum mechanics, shrinks physical standards of distance and time. The math in this scalar theory is much simpler than in GR. For all relativity tests so far, such as the deflection of starlight, the Shapiro time delay, and the rate of precession of the perihelion of Mercury, each theory gives correct answers. The Stanford gpb experiment, due to report 4/2007, will choose between these theories by measuring whether or not there is a spin-orbit coupling between the axis of a nearly perfect gyroscope, in low polar orbit, and the rotating earth beneath it. GR says yes, and MMR says no. Time will tell.

MMR finds that the gravitational influence on the measured metric of space is isotropic, and its consequence for light is equivalent to imbuing space with a scalar index of refraction, "n". Index of refraction is the ratio of the speed of light in a vacuum to the speed in a transparent medium such as glass. Optically measured distance exceeds geometric distance by "n". MMR finds this gravitational "n" to be:

$$n = 1/\alpha^2 \quad \text{where} \quad 1/\alpha = 1 - U/c^2 \qquad [2]$$

U is the scalar gravitational potential, and is negative. This formula works well when gravity is small, but an alternate form is needed when gravity is large:

$$1/\alpha = \exp(-U/c^2) \qquad n = 1/\alpha^2 = \exp(-2U/c^2) \qquad [3]$$

Why not use GR to investigate the gravitational changes of the metric? A compelling reason is that U is almost uniform, and GR is only concerned with the gradient of U. The Einstein equivalence principle underlies GR, equating the effects of gravity on the metric with acceleration. The gravitational acceleration of a test mass is the gradient of U. MMR is directly sensitive to the scalar gravitational potential, instead of to its gradient. GR describes an anisotropic gravitational actual distortion of space along a radius from a single spherical mass, where the gravitational potential at distance r is

$$U = -GM/r \qquad [4]$$

The GR change of the metric along a radius, as for example in the Schwarzschild metric outside a non-rotating spherical mass, finds that distance along a radius is expanded by the factor $1/(1+U)$. This is strange, that the GR metric appears to respond to gravitational potential yet finds no change of the metric when U is uniform like that existing within a hollow massive sphere. No matter how large U may be!

This means that GR is not helpful when the gravitational potential arises from multiple masses. At any point within our universe, we need to account for the gravitational potential arising from billions of galaxies each of which contains billions of stars. The universe is complex, but can be modeled as a sphere of uniform mass density (over a large scale) and with radius $R=cT$ at present time T. It is then possible to calculate U at any radius, $r<R$. [1] We can then deal with the gravitational changes of the metric, using MMR, and find how this modifies the course of light by using the index of refraction [3]. This technique has been used to account for the curvature of Hubble plots (5), in an earlier paper before it was found possible to separate the Doppler shift from the measured red shift.

It will be shown, next, how one can extract the Doppler velocity component, $\beta 1$, from the total red shift, Z. In the absence of acceleration, one expects to find geometric distance proportional to $\beta 1$ per Hubble. We can calculate the optically expanded distance for comparison with the measured distance from a SN1a event that exploded at T* by adding, to the Hubble expectation which is $\beta 1 x T*$ for each data point, the difference in the integral of distance over the light path using (a) the actual "n" along the path and (b) the fixed "n" at the observer.

One should take notice that the measured distance is the distance to where the event occurred. This distance is not the present distance to the remnants of that event. A proper Hubble plot should be of present distance vs. recession velocity, which means that all distances should be extrapolated to a common time, T. This will be done prior to curve fitting the data.

## Separating the Doppler and gravitational red shifts.

The red shift Z is defined by the wavelength increase:

$$Z = (\lambda - \lambda_0)/\lambda_0 \qquad [5]$$

At low velocities, absent gravity, Z is approximately equal to β in the Doppler velocity, βc. At higher velocities, Z exceeds β but can be converted to β using SR (special relativity). The range of β is 0→1, and of Z is 0→∞.

$$\beta = [(Z+1)^2 - 1]/(Z+1)^2 + 1] \qquad [6]$$

Before trying to separate out the Doppler component of Z, β1, prudence suggests we look at the magnitude of the gravitational component to see if separation is needed. There is considerable confusion in the literature about what magnitude we should expect for the gravitational red shift. Some consider the gravitational potential only due to the mass of the star which is the source of the light, and others consider the the mass and size of the galaxy which contains the source. However, the 1/r range of the gravitational potential is so great that one needs to include all contributions within the big bang. The gravitational potential from a single point mass M is -GM/r. To make this dimensionless, we divide by $c^2$. This quantity occurs frequently and will be denoted by $A=GM/rc^2$:

At the surface of earth, neglecting other masses: $A=6.957 \times 10^{-10}$.

At the surface of sun, neglecting other masses: $A=2.119 \times 10^{-6}$.

At the surface of big bang: A=0.3

At the center of big bang: A=0.47

To get the last two numbers, the big bang was modeled with conventional parameters as a sphere of uniform density, of radius $R=cT=14.9 \times 10^9$ light years, and with $M=6 \times 10^{52}$ kg. It is clear that the gravitational potential within the big bang is vastly greater than that we normally would think. We don't pay much attention to it, because it is uniform and so exerts no force on a mass. Because this gravitational potential is inversely proportional to the radius of the big bang, R=cT, the gravitational potential scales as T/t at any prior time t. For example, suppose that a SN1a exploded when the big bang was only half its present size. The wavelength of the emitted light was then only about half what it is now, and the gravitational red shift is Z2=T/(T/2)-1=1. This red shift is comparable with the distant SN1a data of the Riess data set, where red shifts were observed out to Z=1.2, and means we should expect to find that gravity contributes significantly to the observed Z.

## Within our big bang, these two red shifts are interconnected.

We cannot presently see every event that happens in the night sky. For a given precursor of a SN1a event, moving away from us at β1, we at time T can only see its light if it exploded at a specific prior time, T*. If it exploded sooner, its light has already passed us. If it exploded later, that light is still in our future. This time T* is unique to that particular SN1a event, insofar as we are able to see it explode. T* is defined in terms of T and the Doppler velocity, β1. T* is also defined by the gravitational part of the red shift, Z2, which measures the difference in gravitational potentials existing at the observer at T and at the source at T*. This is the common factor that lets us separate the Doppler and gravitational contributions to the measured red shift, Z.

Let a measured red shift Z derive from Z1 due to Doppler velocity and Z2 due to gravity. Let the SN1a precursor move at velocity β1 outward from center for time T*, covering a distance d= β1T*. Let the observer remain at center. The light from the SN1a then travels back to the observer at c in the remaining time, T-T*.

$$d= \beta 1 T^* = T-T^* \quad \text{from which} \quad T^*=T/(1+\beta 1) \quad [7]$$

Gravity also plays a role. The gravitational potential scales as T/T*, and this leads to a red shift, Z2. To get this, we need to know how gravitational potential, U, affects wavelength. When there is no gravitational potential, the wavelength of an atomic emission is designated $\lambda_0$. This is shortened by U, which by [1] and upon neglecting the r/R term is:

$$\lambda = \lambda_0/(1-U/c^2) = \lambda_0/(1+1.5GM/Rc^2) \qquad [8]$$

At least, this is how it goes in weak gravity. The gravitation potential within the big bang, even now, is greatly larger than one might think as was just shown in the comparison of "A" from Earth, from our sun, and within the big bang. It was found necessary to use [3] to modify the form of [8] for very large gravity, to avoid a singularity at small T* when -U exceeds $c^2$.

$$\lambda = \lambda_0 \exp(-1.5GM/Rc^2) \qquad [9]$$

The red shift observed at site 2 from an event that occurred at site 1 is

$$Z = (\lambda 2 - \lambda 1)/\lambda 1 \quad \text{or} \quad Z+1 = \lambda 2/\lambda 1 \qquad [10]$$

$$Z+1 = \lambda_0 \exp(-1.5GM/Rc^2)/\lambda_0 \exp[(-1.5GM/Rc^2)(T/T^*)] \qquad [11]$$

From [7],

$$T/T^* = 1+\beta 1 \qquad [12]$$

and $\beta 1$ can be derived from the Z1 part of Z, attributable to Doppler shift, using [6].

$$\beta 1 = [(Z1+1)^2 - 1]/(Z1+1)^2 + 1] \qquad [13]$$

and so [11] becomes, with $A = GM/Rc^2$,

$$Z+1 = \exp(-1.5A)/\exp[(-1.5A)(T/T^*)] \qquad [14]$$

$$Z+1 = \exp(-1.5A)/\exp[(-1.5A)(1+[(Z1+1)^2 -1]/[(Z1+1)^2+1])] \qquad [15]$$

A second equation is needed, to obtain solutions for Z1 and Z2 for a given Z. Z is not simply the sum of Z1 (due to Doppler velocity) and Z2 (due to gravity). Z+1 is the red shift multiplier that changes $\lambda 1$ to $\lambda 2$. The Doppler and the gravitational red shifts (each plus 1) are hence multiplicative:

$$Z+1 = (Z1+1)(Z2+1) \qquad [16]$$

We then use the "find" function of Mathcad to solve these equations. Upon entering a value for Z, it returns values of Z1 and Z2. The desired Doppler shift is, from [13],

$$\beta 1 = [(Z1+1)^2 -1]/(Z1+1)^2+1] \qquad [17]$$

The Mathcad worksheet is shown next, for Z=1.2. The solutions are: Z1=.502, Z2=.465, and β1=.386.

$$A := 0.66 \qquad Z := 1.2 \qquad Z1 := .1 \qquad Z2 := .2$$

given

$$Z2 + 1 = \frac{e^{1-1.5 \cdot A}}{e^{\left[1 - 1.5 \cdot A \cdot \left[1 + \frac{(Z1+1)^2 - 1}{(Z1+1)^2 + 1}\right]\right]}}$$

$$Z + 1 = (Z1 + 1) \cdot (Z2 + 1)$$

$$\text{Find}(Z1, Z2) = \begin{pmatrix} 0.502 \\ 0.465 \end{pmatrix}$$

$$L := .502$$

$$\beta 1 := \frac{(L+1)^2 - 1}{(L+1)^2 + 1} \qquad \beta 1 = 0.386$$

Table 1. Mathcad worksheet for separation of Z1 and Z2, and calculation of the Doppler shift, β1.

Notice that we needed a trial value for A, and this will eventually be tested by the curve-fitting routine which returns an "A" that fits the data. If these values of A do not agree, one uses a new trial value for A for the separation of the red shift contributions until the curve-fit returns that value of A. The value of A that works for this data set is A=0.66, and the separation of Z into Z1, Z2, and β1 is shown next.

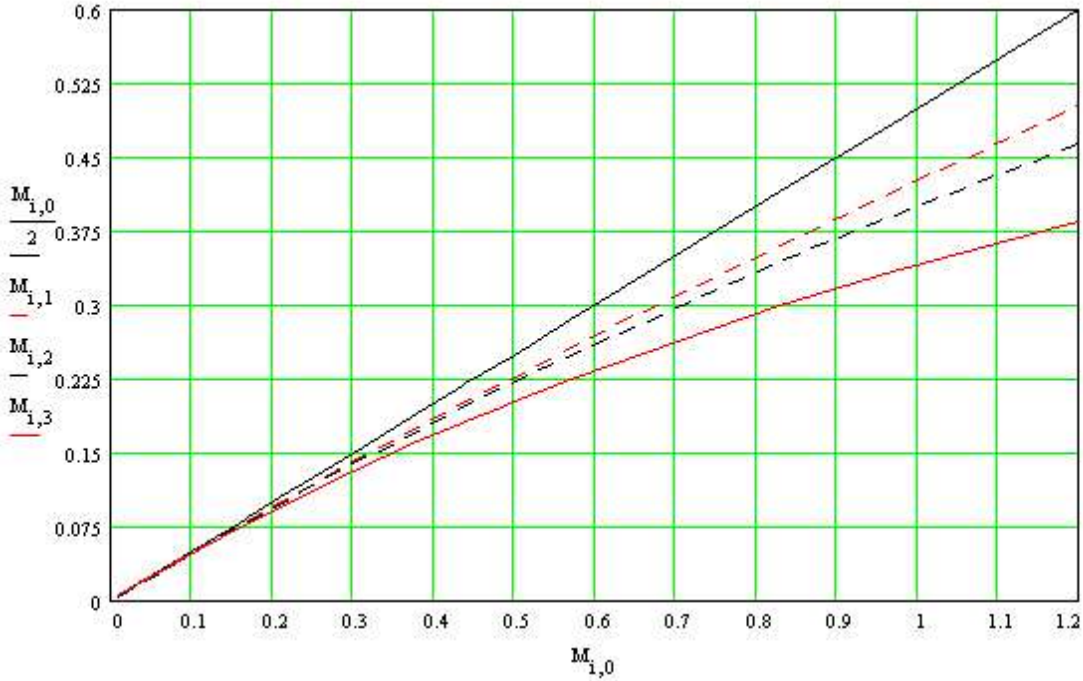

Figure 3. Separation of the red shift components.

Z is plotted along the x axis. Half Z is the straight black line, for reference. Z1 is the red dashed line. Z2 is the black dashed line. The Doppler shift β1 is the red line. It is obvious that one must know "A" in order to separate Z into its components. What value of A will return equal contributions at small red shift? A=.6605. The value of A that fits the data is, coincidentally, A=0.66. The gravitational potential slightly exceeds the Doppler contribution to the red shift, at high red shifts.

## The Hubble plot.

Many direct measurements of the Hubble constant have been made, and the results have been trending towards about $H_0 = 71$ +/- 4 km/s/mpc. Since these numbers were all obtained by assuming that the red shift Z is entirely caused by Doppler shift, and since the gravitational part of the red shift has been shown to be comparable with the Doppler component, it is clear that $H_0$ is smaller, and the universe is older, than has hitherto been thought.

At the large distances afforded by the SN1a supernovae as standard candles, a Hubble plot clearly curves upwards. See Figures 1 and 2, plotted as optically measured distance vs. total red shift Z. These distances, all measured at constant time (now), are from events that occurred at different times. No corrections have been applied to extrapolate to present distances. The conventional interpretation has been that this measured distance is actually geometric distance and is properly measured by optical dimness of a standard candle, the SN1a supernova. The expected distance, absent acceleration, is the blue straight line for which is $H_0 = 70$. Notice that the plot in Figure 2 is approximately linear, out to RS=0.3. When one plots distance vs. Doppler shift, instead of Z, one finds these plots curve upwards even at low red shifts. This curvature at low red shift is quashed, when distance is plotted vs. red shift, by the non-linear nature of the red shift.

## Curve fitting the Hubble data.

With these corrections, we can curve fit the Riess data set and extract information about our universe. The optical distance at time T*, dA, is found by calculating the distance using the variable index of refraction over the path, less the distance calculated using the fixed index of refraction at the observer, plus the Hubble expectation that geometric distance is velocity times T*. This distance dA is then extrapolated to the present time T by multiplying by 1+β1. Observed distances are shown by red boxes, and have been multiplied by 1+β1 in order to extrapolate to present distance at time T. The actual formulae were evaluated using Mathcad, and are shown next. The possibility that we are offset from center is had by introducing the parameter, "a", the fraction of R that we are offset from center. "dn" is the difference between optical distance and geometric distance. T* is T/(1+β1).

The curve fitting protocol is rudimentary. At first, a=0 and different choices of A are made for $T=13.8 \times 10^9$ yr corresponding to the consensus Hubble constant of 71. The standard deviation is the measure of the fit, and the SD plotted vs. different values of A is a parabola, concave upwards. Choosing the A that gives lowest SD, T is then varied for best fit. Finally, "a" is allowed to vary until best fit is found. The process is then repeated, with new trial values using the parameters that gave the best fit. If more computational power were available, one could include the division of the red shift in the curve fit routine and rapidly extract the desired solutions.

$$A := 0.6674 \quad T := 23.5 \cdot 10^9 \quad a := 0.13 \quad\quad R_i := \beta 1_i \cdot T$$

$$dA_i := \int_{\frac{T}{1+\beta 1_i}}^{T} e^{3 \cdot A \cdot \frac{T}{t} \cdot \left[1 - \frac{1}{3}\left[\left(\frac{T-t}{T}\right) + a\right]^2\right]} dt - \int_{\frac{T}{1+\beta 1_i}}^{T} e^{3 \cdot A \cdot \left(1 - \frac{a^2}{3}\right)} dt + \beta 1_i \cdot \frac{T}{1+\beta 1_i}$$

$$dA_i := (1 + \beta 1_i) \cdot dA_i$$

$$dn_i := dA_i - \beta 1_i \cdot T \quad\quad SD := \sqrt{\frac{\sum_i \left[(dA_i) - r_i\right]^2}{N}} \quad\quad R_i := \beta 1_i \cdot T$$

$$SD = 7.659 \cdot 10^8$$

Table 2. Curve-fitting formulas evaluated using Mathcad.

    The curve fit to the data points is shown next, Figure 4. As before, the data points are shown as red boxes. The curve fit is the blue line. The dashed red line shows the increase of optical distance over geometrical distance in the metric of the observer (us). The black line is the Hubble expectation, i.e. the geometrical distance, also in the metric of the observer. The parameters, varied to get best fit, are A, T, and a. A is $GM/Rc^2$, T is the time since the big bang, and "a" is our radial fractional distance from center. Not surprisingly, the actual values of these parameters differ somewhat from the prior consensus findings. Using conventional values for $M=6 \times 10^{52}$ kg and $T = 13.8 \times 10^9$ years (T=R for c=1), one expects to find A=.34. The found value of A=0.6674 corresponds to a total mass of the universe of $M = 21.1 \times 10^{52}$ kg, when combined with the larger found value of T. The age T found in this study is larger than has been thought, $23.5 \times 10^9$ years instead of $13.8 \times 10^9$ years. This shrinks the Hubble constant, from 71 to 41.6. Several tries were needed to get the "found" value of A to be approximately equal to the A used in [15] to divide the Doppler and gravitational components of the red shift.

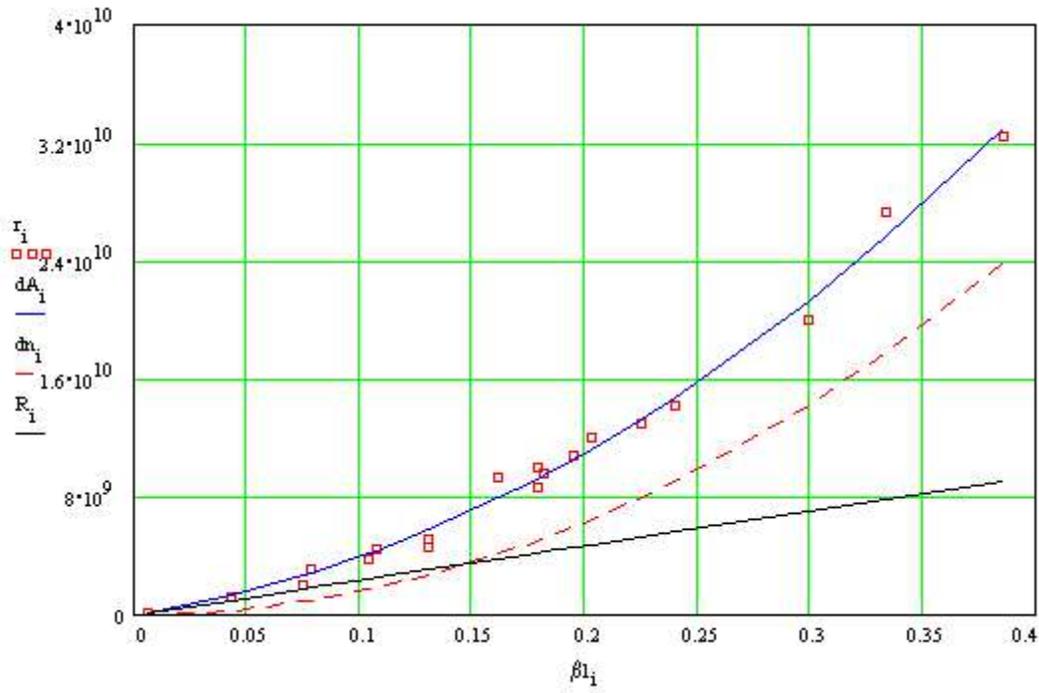

Figure 4. Curve fit of measured distance vs. Doppler velocity.

A proper Hubble plot should be of actual present distance vs. actual Doppler velocity. The next figure subtracts, from the extrapolated observed distances, the optical effects of the metric, yielding the extrapolated geometric distance at common time, T. This is the purest implementation of the Hubble concept.

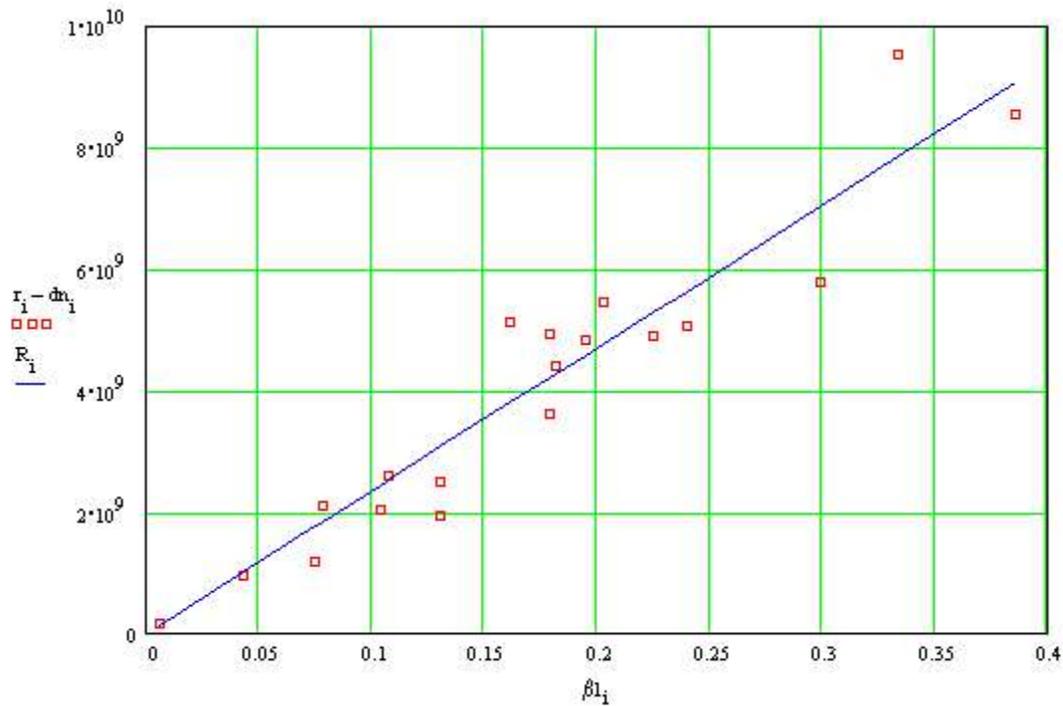

Figure 5.  Corrected Hubble plot, geometric distance vs. Doppler velocity.

The Hubble plot of Figure 5 is not as pretty as that shown in Figure 4, since  the "optical" contributions to distance have been subtracted and the noise level is then more apparent on the smaller scale of distance.

As for "a", our radial distance from center, the data is best fitted by an "a" of 13%  of the radius of the big bang but the precision of the data is not great.  I'm unaware of any prior suggestions of where we lie within our universe, and this number "a" will probably change as more and better data become available. Even if the precision of the data persuades one that this value of "a" is meaningful, we don't really know that we are looking radially outwards.  Several attempts to curve fit Riess' data based on "us" being outside and the SN1a sources lying close to center were, however, less successful.  To the extent that one considers this "a" meaningful, he should consider it a minimum since the variation of the gravitational potential over the light path is greatest when looking along a radius.  One hopes the SNAP Hubble replacement telescope will include data taken from different directions than the present data set from the Hubble deep field north.  Such data should be available within a decade, assuming the replacement space telescope launches successfully in a few years. The greater precision expected of the Hubble data from this probe will be most welcome.

## Discussion.

The cosmological principle precludes that we should ever find any structure in our universe, other than locally. It holds that, over a large enough scale, we must expect to find the same things in every direction. Further, it holds that the distribution of those "things" should be uniform. In the model used here, the variation of gravitational potential within the big bang is a basis for distinguishing where within the big bang we presently exist. As such, it denies the cosmological principle.

What is the theoretical limit to seeing, i.e. the optical horizon? The maximum Doppler shift is $\beta_1=1$, in which case the limit of seeing is $R^*=cT^* =R/2$ because we must allow time for the light to return to us. Another way of looking at it is to recognize that anything that happens at the outer edge of our big bang is not immediately seen by an observer near center. If the event happens at a time $T^*$, and if the edge is receding at c, it will be observed at $2T^*$. The observer sees events as they were at $T^*$, but the edge is by then twice as far away. At best, we can visually explore only 1/8 of the volume of our universe. No matter how large is the gravitational contribution to the red shift, it is the Doppler shift at $\beta_1=1$ that makes the wavelength become infinite and hence unobservable. The metric distortion imposes an additional burden on telescope technology, in that greater sensitivity is required than what might have been thought. This is shown in the next figure. The real or geometric distance, in the metric of the observer, is shown as the blue line, and the optically measured distance is shown as the red line. This means that far measurements seem about 4 times as far as they actually are, and are about 16 times dimmer than one might expect using the inverse square law of illumination.

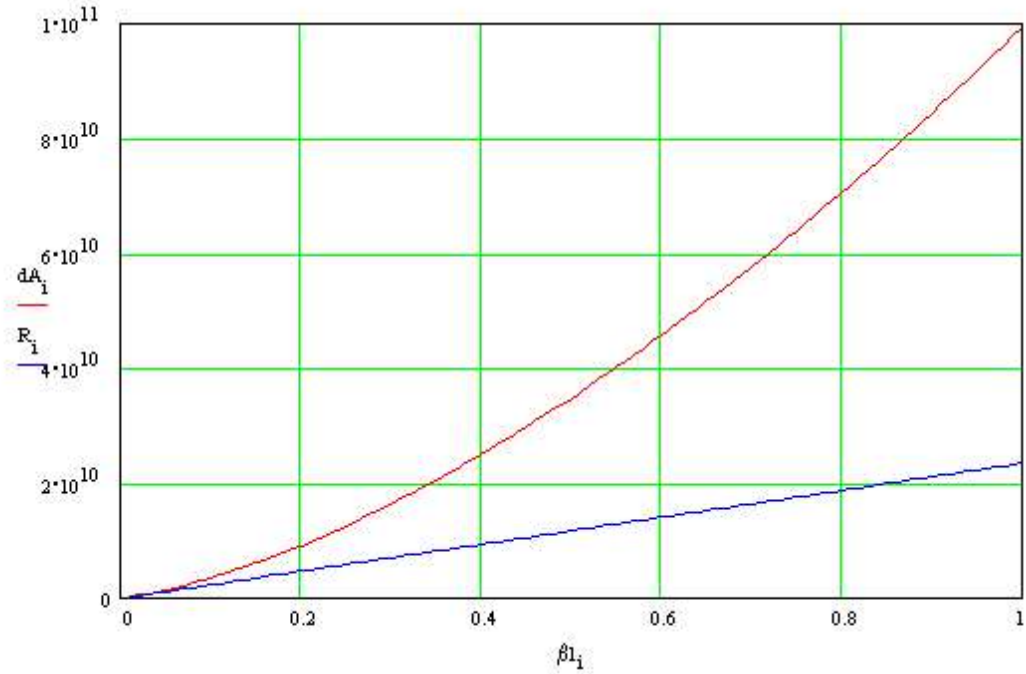

Figure 6. Metric distortion of distance. Optical distance red line, geometrical distance blue line.

Fortunately, these optical measurements of distant objects are measurably influenced by the gravitational potential, a field of much longer range ($1/r$) than that of light which is inverse square ($1/r^2$). Within our optical range, light is affected by the gravitational potential throughout its long journey to us and this extends our sensor range to include all the mass within our universe. This means that we can sense further than we can see! We can derive information from places we can never see, beyond our optical horizon. As we come to understand more about how this very large, less-than-uniform, and time-dependent gravitational potential can be sensed, and as more and better data become available, there is hope that we may finally locate ourselves within our big bang. And, if the mass density is not uniform over the large scale, that too may become known.

The measured value of the Hubble constant has shrunk markedly, since Hubble's early result of 500 km/s/mpc in 1929. Late in the 20$^{th}$ century, it developed two peaks (at 50 and 100). The HST (Hubble space telescope) results of 2001 (7) claim to have put to rest these differences, and found $H_0 = 72(\pm 4)$. An indirect method of obtaining $H_0$ without use of a Hubble plot was reported by Lineweaver and Barbosa (8), who relied on adiabatic expansion of the CMB (cosmic microwave background) and prefer a value of 30 (-7, +18). They conclude

".....current CMB data favor a low value for the Hubble constant. Such low $H_0$ models are consistent with Big Bang nucleosynthesis, cluster baryonic fractions, the large-scale distribution of galaxies and the ages of globular clusters; *although in disagreement with direct determinations of the Hubble constant*."

(Italics added.) This reference was found after this paper had essentially been written. Their conclusion is of particular interest because it does not rely on the assumption that the red shift is wholly Doppler. The same group later narrowed it down to $H_0 = 35$ (9), which is closer to the finding of this study, $H_0 = 41.6$.

## Conclusions.

The Hubble data now available from events out to red shift Z=1.2 has inflamed the imagination of many scientists. The unexpected curvature of these plots of observed distance vs. red shift has led to strange ideas, with talk of an accelerating or runaway universe somehow powered by an unknown "dark energy". An easier explanation is at hand, based on the gravitational changes of the metric and assisted by finding a way to separate the Doppler and gravitational contributions to the measured red shift. General relativity has been the standard for dealing with gravity, since Einstein introduced it some 80 years ago. MMR was introduced only 5 years ago, and is not yet generally accepted for events where MMR and GR predict different outcomes. An important test of these relativity theories is now running, the Stanford gpb. This apparently successful project is winding down, and final results are expected in April 2007. The gpb will either accept or reject the GR claim that gravity distorts the very metric of space, asymmetrically, which must lead to a Lense-Thirring gravitational spin-orbit coupling between a gyroscope in low polar orbit with the spinning earth beneath. The gpb results will clearly deny at least one of these theories, GR or MMR.

The high Z Hubble data offer an interesting test for relativistic theories of gravity. One finds that MMR can cope with gravitational potential, while GR cannot. That is, GR finds no gravitational influence on the metric in the presence of a large and uniform gravitational potential. For example, GR finds no gravitational distortion of the metric within a massive hollow sphere. If the use of MMR to explain the curvature of the Hubble plot were incorrect, it seems unlikely that that its application should lead to a value of the total mass of the universe comparable to that derived from other analyses.

Cosmology is a much studied subject, but mostly seems to accept that the metric of the universe is and always has been just like our present metric here on Earth. It calculates, for example, the total gravitational energy of all the mass at its present distance and then worries about whether the net energy, kinetic plus potential, is positive or negative. If the argument is persuasive, that the metric of space depends on the gravitational potential and that this changes with time, the cosmological calculations are inherently unfounded. The single exception lies with "inflation" theory, based on GR and introduced to explain how things can move faster than light in the early stages of the big bang. The curved Hubble data plots require an explanation, and find it in the changes of the metric arising from the gravitational potential derived from all the mass in our universe. The simplistic assumption that red shift is wholly Doppler, for these objects in the night sky, has led to a systematic error that, since 1929, has more or less doubled the number obtained for the Hubble constant from a Hubble plot. With the separation of the Doppler shift from the red shift, the Hubble constant shrinks from 71 to 41.6. As for the mass density, which has been thought to control the expansion rate of the universe, these new values of M and R imply a slight reduction of the mass density to 71% of the prior consensus value.

# References.